\documentclass[10pt,pra,twocolumn,superscriptaddress,showpacs,floatfix]{revtex4}

\usepackage{graphicx}
\usepackage{amssymb}
\usepackage{times}
\usepackage{amsmath}
\usepackage{amsthm}
\usepackage{epstopdf}

\usepackage{hyperref}

\input epsf.tex
\usepackage{graphicx}
\usepackage{amsthm}
\usepackage{subfigure}

\begin{document}
\title{Practical security of continuous-variable quantum key distribution with reduced optical attenuation}% Force line breaks with\\
\author{Yi Zheng}
\affiliation
 {College of Information Science and Technology, Northwest University, Xi'an 710127, Shaanxi, China
}
\author{Peng Huang}\thanks{Corresponding author: huang.peng@sjtu.edu.cn}
\affiliation
 {State Key Laboratory of Advanced Optical Communication Systems and Networks, and Center of Quantum Information Sensing and Processing, Shanghai Jiao Tong University, Shanghai 200240, China
}
\author{Anqi Huang}
\affiliation
 {Institute for Quantum Information $\&$ State Key Laboratory of High Performance Computing, College of Computer, National University of Defense Technology, Changsha 410073, China
 }
\affiliation
 {Greatwall Quantum Laboratory, China Greatwall Technology, Changsha 410205, China
 }
\author{Jinye Peng}
\affiliation
{College of Information Science and Technology, Northwest University, Xi'an 710127, Shaanxi, China
}
\author{Guihua Zeng}\thanks{Corresponding author: ghzeng@sjtu.edu.cn}
\affiliation
 {State Key Laboratory of Advanced Optical Communication Systems and Networks, and Center of Quantum Information Sensing and Processing, Shanghai Jiao Tong University, Shanghai 200240, China
}
%\date{\today}
\begin{abstract}
In a practical CVQKD system, the optical attenuator can adjust the Gaussian-modulated coherent states and the local oscillator signal to an optimal value for guaranteeing the security of the system and optimizing the performance of the system. However, the performance of the optical attenuator may deteriorate due to the intentional and unintentional damage of the device. In this paper, we investigate the practical security of a CVQKD system with reduced optical attenuation. We find that the secret key rate of the system may be overestimated based on the investigation of parameter estimation under the effects of reduced optical attenuation. This opens a security loophole for Eve to successfully perform an intercept-resend attack in a practical CVQKD system. To close this loophole, we add an optical fuse at Alice's output port and design a scheme to monitor the level of optical attenuation in real time, which can make the secret key rate of the system evaluated precisely. The analysis shows that these countermeasures can effectively resist this potential attack.
\end{abstract}
%\textbf{Keywords:} plug-and-play, continuous variable, measurement-device-independent, quantum key distribution, noiseless linear amplifier.
\pacs{03.67.Hk, 03.67.-a, 03.67.Dd}
\maketitle
\section{INTRODUCTION}\label{sec1}
In the field of quantum cryptography, quantum key distribution (QKD) technology has attracted widely attention due to its unconditional security based on the guarantee of the basic laws of quantum physics \cite{Ekert1991Quantum,Lo1999Unconditional}. It is a relatively mature technology to share symmetric keys between the authorized communication parties Alice and Bob through an insecure quantum channel that can be freely controlled and manipulated by a potential eavesdroppers Eve \cite{Ekert1991Quantum,Lo1999Unconditional,gisin2002quantum,weedbrook2012gaussian}. At present, QKD systems can be divided into the two broad classes: discrete-variable quantum key distribution (DVQKD) and continuous-variable quantum key distribution (CVQKD). In comparison with DVQKD systems, CVQKD systems using weak coherent states and a homodyne detector can be well compatible with the existing optical communication systems \cite{weedbrook2012gaussian,Takesue2007Quantum,grosshans2003quantum,qi2007experimental,fossier2009improvement}. Therefore, research of CVQKD has far-reaching significance.

In particular, CVQKD with the Gaussian-modulated coherent states (GMCS) is the most favorable scheme, which has been experimentally implemented by many research groups in laboratories and in field environment \cite{qi2007experimental,fossier2009field,Jouguet2012Experimental,huang2016field,wang201525}. Here, we also focus on the exploration of GMCS CVQKD schemes. Although GMCS CVQKD schemes have been theoretically proven to be secure against the collective and coherent attacks \cite{Jouguet2012Analysis,leverrier2013security}, there are some security loopholes that can be used for hiding their attacks by Eve in practical experimentations of such schemes, such as the local oscillator (LO) fluctuation attack \cite{Ma2014Local}, the LO calibration attack \cite{Jouguet2013Preventing}, the wavelength attack \cite{Huang2013Quantum,Ma2014Wavelength}, the saturation attack \cite{Qin2016Quantum}, finite sampling bandwidth effects \cite{Wang2016Practical}, homodyne detector blinding attack \cite{qin2018homodyne}, jitter in clock synchronization \cite{xie2018practical}, and the polarization attack \cite{zhao2018polarization}. Subsequently, many corresponding countermeasures have been proposed by researchers. To be specific, in order to resist the attacks related to LO, a real-time shot-noise measurement (RTSNM) scheme is proposed by researchers \cite{Kunzjacques2015Robust,Wang2018Practical}. Recently, a local LO (LLO) CVQKD scheme is designed and implemented experimentally, which can fundamentally close the security loophole originates from LO \cite{huang2015high-speed,soh2015self,qi2015generating,tao2018high,wang2018pilot}. In addition, the proposed continuous-variable measurement-device-independent quantum key distribution (CV-MDI-QKD) schemes can completely eradicate all fatal loopholes related to the detector \cite{pirandola2015high-rate,li2014continuous-variable}. This history shows that the discovery of security loopholes promotes study of corresponding countermeasures, enhancing the practical security of CVQKD. Therefore, the explorations of the concealed security loopholes are vital to the practical application of CVQKD.

Optical attenuation is one of the key steps in the implementation of CVQKD schemes. However, the performance of optical attenuators may deteriorate due to environmental perturbations. For example, Eve can reduce its attenuation level through laser damage attack \cite{huang2018quantum,bugge2014laser}. In particular, the influences, which originate from the decrease of optical attenuation, are ambiguous for the performance of a practical CVQKD system. In this paper, we mainly investigate how the decrease of optical attenuation affects the performance of the system. More specifically, we first reveal the deviation between the ideal states and the transmitted Gaussian-modulated states affected by the reduced optical attenuation. Then, we explore the quantum channel parameter estimation under these effects based on the above analysis. Our results show that the channel excess noise may be underestimated by Alice and Bob under the effects of decreased optical attenuation. Subsequently, the secret key rate is overestimated by Alice and Bob, which is coincident with the security analysis result of DVQKD under the effects of reduced optical attenuation \cite{huang2019laser}. Thus, the decrease of optical attenuation may open a security loophole for Eve to successfully hide her attacks, such as the classical non-Gaussian attack, i.e., the intercept-resend attack. To eliminate this loophole, we add an optical fuse at Alice's output port and design a real-time monitoring scheme of the level of optical attenuation for a practical CVQKD system. The analysis indicates that Alice and Bob can precisely evaluate the channel parameters to accurately calculate the secret key rate for a practical CVQKD system through this scheme.

This paper is organized as follows. In Sec. \ref{sec2}, the effects of decreased optical attenuation are described and modeled. Then, we analyse the parameter estimation of GMCS CVQKD systems with reduced optical attenuation in Sec. \ref{sec3}. By considering the effects of reduced optical attenuation, in Sec. \ref{sec4}, we evaluate the secret key rate against collective attacks through theoretical calculation and simulation. To resist the loophole induced by decreased optical attenuation, we add an optical fuse at Alice's output port and design a real-time monitoring method for optical attenuation in Sec. \ref{sec5}. Finally, conclusions are presented in Sec. \ref{sec6}.
\section{THE EFFECTS OF REDUCED OPTICAL ATTENUATION}\label{sec2}
\begin{figure}[!h]\center
\centering
\resizebox{8.5cm}{!}{
\includegraphics{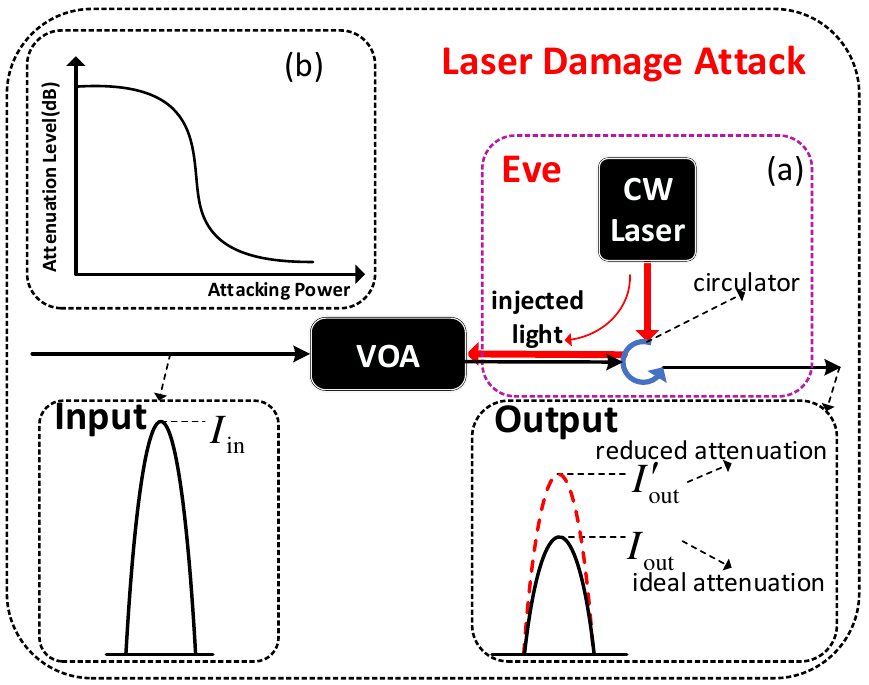}}
\caption{The performance of VOA under the laser damage attack. Part (a) describes the schematic of the laser damage attack. Part (b) shows the relation between the attenuation level of VOA and the attacking power. CW Laser, continuous-wave laser; $I^{\prime}_{\text{out}}$, the intensity of the output optical signal with reduced attenuation.}
\label{FIG1}
\end{figure}

Optical attenuators can reduce the intensity of their input optical signals to satisfy with the request of QKD systems. In particular, variable optical attenuator (VOA) is one well-known attenuator that can adjust the level of attenuation. The level of attenuation is represented as
\begin{equation}
\label{eq1}
L=-10\lg \frac{I_{\text{out}}}{I_{\text{in}}},
\end{equation}
where $I_{\text{in}}$ and $I_{\text{out}}$  are the intensity of the input and output optical signals in VOA, respectively. However, in practical experimentation, the level of attenuation may not achieve the excepted goal due to the effects of environmental perturbations. For example, Eve may perform the laser damage attack on VOA to reduce its attenuation level in a running DVQKD system. Specifically, Eve can inject suitable light to the VOA via quantum channel to deteriorate the performance of the attenuator (see Fig. \ref{FIG1}) \cite{huang2018quantum,bugge2014laser}. In particular, the VOA can suffer catastrophic damage when the attack reaches a certain power, which has been shown experimentally \cite{huang2018quantum}. Fig. \ref{FIG1} also reflects the relationship between the intensity of the output and input optical signal in VOA under the effects of the laser damage attack based on the Eq.(\ref{eq1}). Obviously, the intensity of the output optical signal for the attacked VOA is greater than its ideal value.
\begin{figure}[!h]\center
\centering
\resizebox{8.5cm}{!}{
\includegraphics{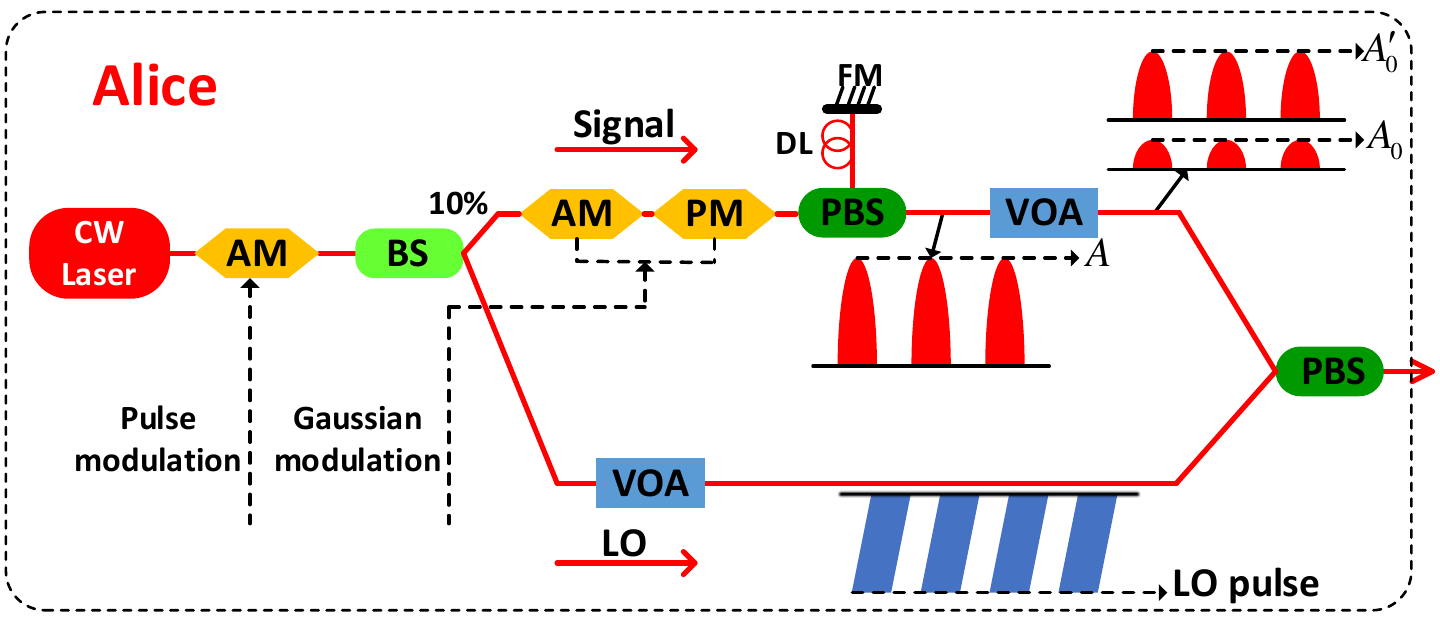}}
\caption{The transmitter of a general practical CVQKD system. AM, amplitude modulator; PM, phase modulator; BS, beam splitter; PBS, polarizing beam splitter; FM, faraday mirror; DL, delay line; LO, local oscillator.}
\label{FIG2}
\end{figure}

Optical attenuator is also a key device for the implementation of a CVQKD system. Similarly, VOA has been widely used in the CVQKD systems. In particular, VOA is used in the transmitter of a practical CVQKD system (see Fig. \ref{FIG2}), which can attenuate the intensity of the Gaussian-modulated optical signal $A$ to quantum level for guaranteeing the unconditional security of the system based on the basic laws of quantum physics \cite{grosshans2003quantum,fossier2009improvement,leverrier2013security,weedbrook2012gaussian}. It is notable that the attenuation level of VOA in the CVQKD systems can be also decreased by Eve through laser damage attack. Here, after attenuation, we use $A_0$ and $A^\prime_0$ to represent the ideal quantum signal and imperfect quantum signal, respectively.

More specifically, in a Gaussian-modulated scenario, Alice generates a series of Gaussian-modulated coherent states $|\alpha_A\rangle$ based on the key information \cite{weedbrook2012gaussian}. In the phase space, $|\alpha_A\rangle$ can be expressed as
\begin{equation}
\label{eq2}
\begin{split}
&|\alpha_A\rangle=|\alpha_A|e^{i\theta}=x_A+ip_A,\\
&x_A=|\alpha_A|\cos\theta,p_A=|\alpha_A|\sin\theta,
\end{split}
\end{equation}
where $|\alpha_A|$ and $\theta$ respectively indicate the amplitude and phase of the Gaussian-modulated optical signal $A$, $x_A$ and $p_A$ are two independent quadratures variables with identical variance $V_A$ and zero mean \cite{grosshans2003quantum,weedbrook2012gaussian}. Moreover, the intensity $I_A$ and amplitude $|\alpha_A|$ of the optical signal $A$ obey the following relation:
\begin{equation}
\label{eq3}
I_A\propto {|\alpha_A|}^2
\end{equation}
Therefore, Alice can adjust the values of $x_A$ and $p_A$ to the appropriate values $x_{A_0}$ and $p_{A_0}$ by VOA. Here, $x_{A_0}$ and $p_{A_0}$ are two quadratures variables of the ideal quantum optical signal $A_0$. Correspondingly, $|\alpha_{A_0}\rangle=x_{A_0}+ip_{A_0}$ is the transmitted Gaussian-modulated coherent states. In addition, the variance $V_A$ of the quadratures variables $x_A$ or $p_A$ in terms of the mean photon number of optical signal $A$ reads \cite{laudenbach2018continuous}
\begin{equation}
\label{eq4}
V_A=2\langle n \rangle.
\end{equation}
Equally, the variance $V_A$ can also be attenuated to the preset value $V_{A_0}$ by Alice through VOA in a practical CVQKD system.
\begin{figure}[!h]\center
\centering
\resizebox{6.5cm}{!}{
\includegraphics{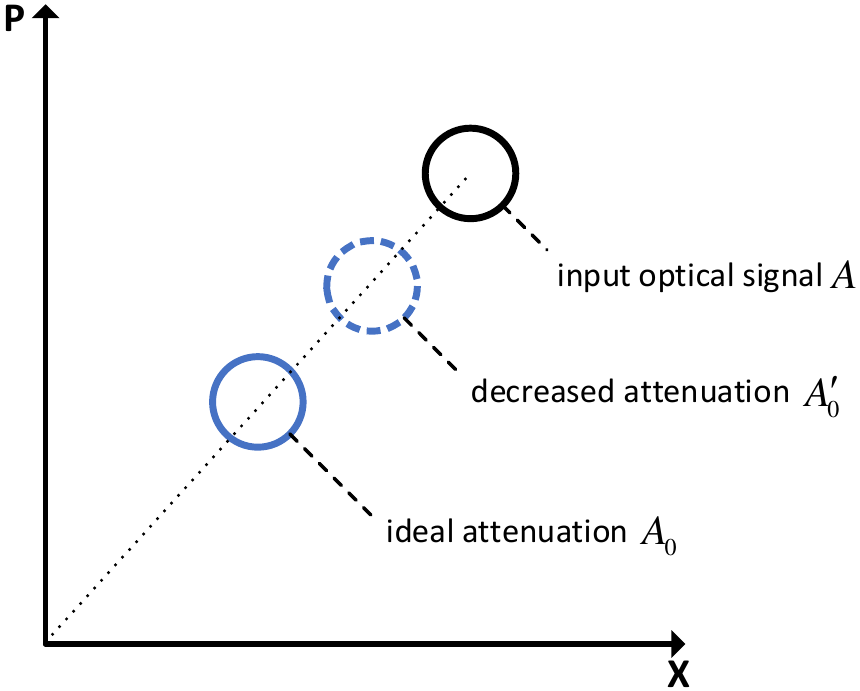}}
\caption{The expression of the transmitted Gaussian-modulated coherent states in the phase space under the effects of reduced optical attenuation.}
\label{FIG3}
\end{figure}

However, $x_{A_0}$, $p_{A_0}$ and $V_{A_0}$ may deviate from the ideal values due to the effects of reduced optical attenuation, which is revealed in Fig. \ref{FIG3} based on the phase space. In the following analysis, we assume that $I^\prime_{\text{out}}$ and $I_{\text{out}}$ satisfy with the following relation:
\begin{equation}
\label{eq5}
I^\prime_{\text{out}}=k I_{\text{out}}(k>1).
\end{equation}
Correspondingly, according to the above analysis, the changes of $x_{A_0}$, $p_{A_0}$ and $V_{A_0}$ are as follows:
\begin{equation}
\label{eq6}
\begin{split}
&x^\prime_{A_0}=\sqrt{k}x_{A_0},p^\prime_{A_0}=\sqrt{k}p_{A_0},\\
&V^\prime_{A_0}=kV_{A_0},
\end{split}
\end{equation}
where $x^\prime_{A_0}$ and $p^\prime_{A_0}$ are two independent quadratures variables of imperfect quantum signal $A^\prime_0$, $V^\prime_{A_0}$ are the variance of $x^\prime_{A_0}$ or $p^\prime_{A_0}$.

\begin{figure}[!h]\center
\centering
\resizebox{8.5cm}{!}{
\includegraphics{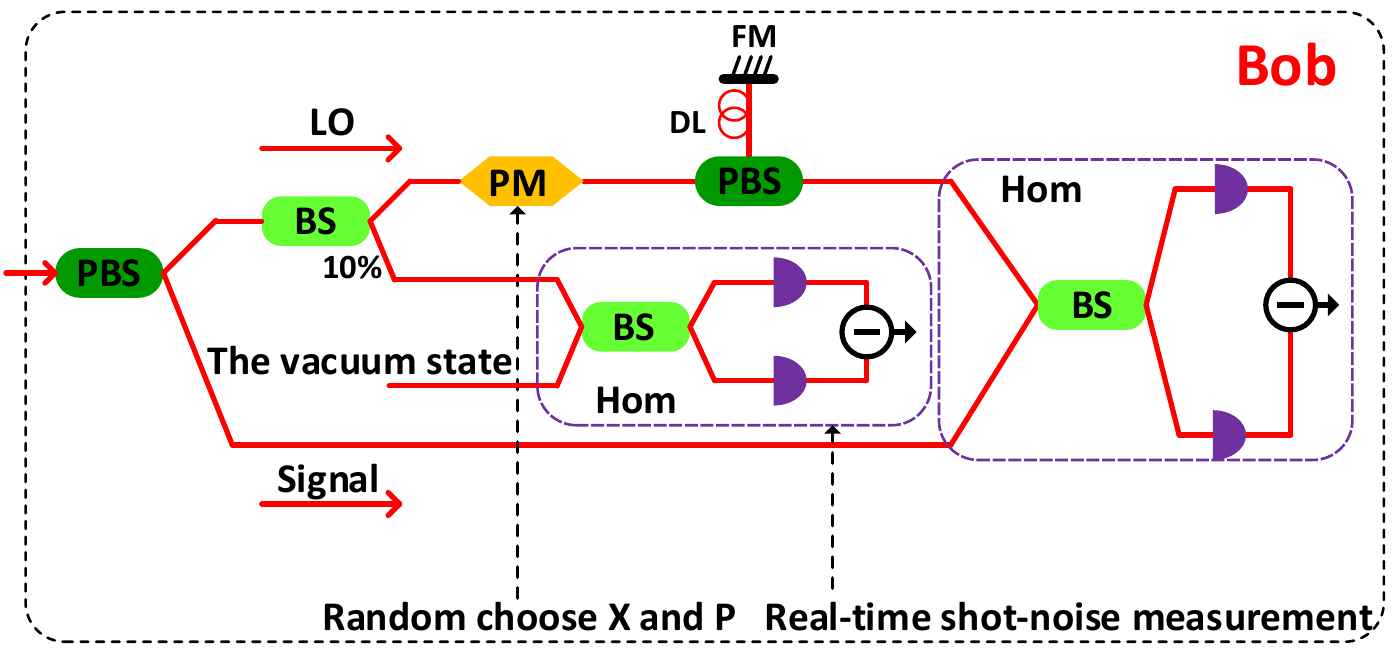}}
\caption{The scheme of the real-time shot-noise measurement in the receiver of a general practical CVQKD system. Hom, homodyne detector.}
\label{FIG4}
\end{figure}
Apart from these, the intensity of LO signal needs to be adjusted to an optimal value by VOA to optimize the performance of a practical CVQKD system. There is no doubt that the intensity of LO signal can deviate from the optimal value under the effects of reduced optical attenuation. In practice, however, the effects of decreased optical attenuation on the LO signal can be completely eliminated through the real-time shot-noise measurement technique. The basic principle of the technique is illustrated in the Fig. \ref{FIG4}. Specifically, Bob first splits a fraction of LO signal to a balanced homodyne detector. Further, the variance of shot-noise can be evaluated through the interference between the separated LO signal and the vacuum mode. In addition, according to the calibrated linear relation between the variance of shot-noise and the intensity of LO signal, the variance can be also evaluated by monitoring the intensity of the LO signal in real time. Accordingly, we will focus on the investigation of the effects of reduced optical attenuation on signal states, i.e., the Gaussian-modulated coherent states, in the following analysis.
\section{PARAMETER ESTIMATION UNDER THE EFFECTS OF REDUCED OPTICAL ATTENUATION}\label{sec3}
\begin{figure}[!h]\center
\centering
\resizebox{8cm}{!}{
\includegraphics{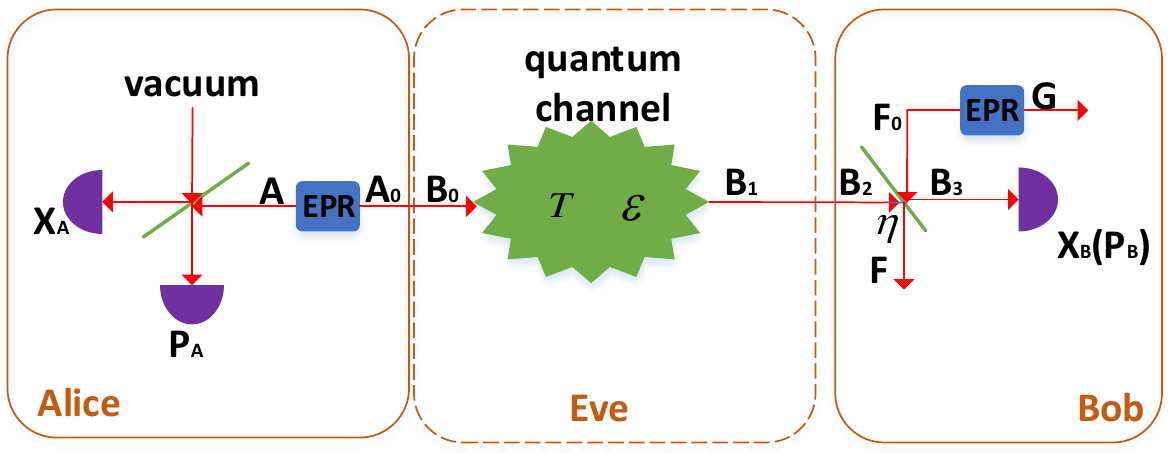}}
\caption{The EB model of the one-way GMCS CVQKD protocols using homodyne detector. The quantum channel parameters, i.e., transmission $T$ and excess noise $\varepsilon$, which can be freely controlled and manipulated by Eve.}
\label{FIG5}
\end{figure}

The key distribution process of the one-way GMCS CVQKD systems can be described by the standard prepare-and-measure (PM) model. However, in practical implementation of CVQKD schemes, the evaluation of the secret key rate is accomplished based on the equivalent entanglement-based (EB) scheme. The sender Alice first prepares one two-mode squeezed (EPR) state with variance $V=V_{A_0}+1$. Subsequently, the mode $A$ is measured by a heterodyne detector, and the mod $A_0$ is sent to the the receiver Bob through a quantum channel with transmissivity $T$ and excess noise $\varepsilon$. Then, Bob exploits a homodyne detector with efficiency $\eta$ and noise $\nu_{\text{el}}$ to measure the quadratures $x_B$ or $p_B$ of his received mode $B_2$ (see Fig. \ref{FIG5}) \cite{fossier2009improvement}. In addition, the homodyne detector inefficiency is modelled by a beam splitter. Finally, it is essential to perform classical post-processing which includes sifting, parameter estimation, key reconciliation and privacy amplification \cite{fossier2009improvement,grosshans2003quantum,Wang2018High,li2018memory}. In particular, parameter estimation is a key step which determines the secret key rate and the maximal transmission distance for a practical CVQKD system.

After transmission and detection of the quantum coherent states, Alice and Bob share a group correlated vectors $X=\{(x_{{A_0}_i},{x_B}_i)|i=1,2,\ldots,N\}$ or $P=\{(p_{{A_0}_i},{p_B}_i)|i=1,2,\ldots,N\}$. Here, $N$ is the total number of the received pulses. The involved quantum channel for GMCS CVQKD systems is assumed to be a normal linear model which satisfies the following relation between Alice and Bob:
\begin{equation}
\label{eq7}
x_B=tx_{A_0}+z,
\end{equation}
where $t=\sqrt{\eta t}$, vector $z$ is the total noises term following a centered normal distribution with variance $\sigma^2=\eta T \xi+N_0+V_{\text{el}}$. Here, the related parameter $\eta$ is detection efficiency of the homodyne detector, $V_{\text{el}}$ is electronic noise of the homodyne detector, $N_0$ is the variance of shot-noise. Accordingly, $x_{A_0}$ and $x_B$ meet the following relations:
\begin{equation}
\label{eq8}
\begin{split}
&\langle {x^2_{A_0}} \rangle=V_{X_{A_0}}, \langle{x_{A_0}}{x_B}\rangle=\sqrt{\eta T}V_{X_{A_0}},\\
&\langle {x^2_B} \rangle=\eta T V_{X_{A_0}}+\eta T \xi+N_0+V_{\text{el}}.
\end{split}
\end{equation}
There is no doubt that $p_{A_0}$ and $p_B$ also satisfy Eqs. (\ref{eq7}) and (\ref{eq8}). Moreover, in the evaluation of the secret key rate, the above parameters $V_{X_{A_0}}$, $\xi$ and $V_{\text{el}}$ must be described by $V_{A_0}N_0$, $\varepsilon N_0$ and $\nu_{\text{el}} N_0$, respectively.

In practical experimentation, Alice and Bob randomly extract $m(m<N)$ pairs data from $X$ or $P$ vectors to estimate the parameters $T$ and $\xi$. The maximum-likelihood estimators $\hat{t}$ and ${\hat\sigma}^2$ are known with the normal linear model \cite{leverrier2010finite,Jouguet2012Analysis}, which are
\begin{equation}
\label{eq9}
\hat t=\frac{\sum_{i=1}^m{x_{{A_0}_i}}{{x_B}_i}}{\sum_{i=1}^m{x^2_{{A_0}_i}}},{\hat\sigma}^2=\frac{1}{m}\displaystyle\sum_{i=1}^m({x_B}_i-\hat t {x^2_{A_0}}_i).
\end{equation}
Here, parameters $\hat{t}$ and ${\hat\sigma}^2$ are independent estimators which obey the distributions
\begin{equation}
\label{eq10}
\hat t\sim \mathcal N \Bigg(t,\frac{\sigma^2}{\sum_{i=1}^m{{x^2_{A_0}}_i}}\Bigg), \frac{m{\hat{\sigma}}^2}{\sigma^2}\sim\chi^2(m-1),
\end{equation}
where $t$ and ${\sigma}^2$ are the true values of these parameters. The $\chi^2$ distribution converges to a normal distribution under the condition that $m$ is large enough (e.g., $m>10^6$). Therefore, the confidence intervals for the estimators $\hat t$ and $\hat\sigma^2$ are
\begin{equation}
\label{eq11}
t\in[\hat t-\Delta t,\hat t+\Delta t], \sigma^2\in[\hat\sigma^2-\Delta \sigma^2,\hat\sigma^2+\Delta\sigma^2].
\end{equation}
Here,
\begin{equation}
\label{eq12}
\Delta t=z_{\epsilon_{\text{PE}}/{2}}\sqrt{\frac{\hat\sigma^2}{mV_{X_{A_0}}}},\Delta \sigma^2=z_{\epsilon_{\text{PE}}/{2}}\frac{\hat\sigma^2\sqrt2}{\sqrt m},
\end{equation}
where $\epsilon_{\text{PE}}$ is a probability that the estimated parameter does not belong to the confidence region, $z_{\epsilon_{\text{PE}}/{2}}$ is a coefficient following $1-\frac{1}{2}\text{erf}(z_{\epsilon_{\text{PE}}/{2}}/\sqrt2)=\frac{1}{2}\epsilon_{\text{PE}}$, and $\text{erf}(\cdot)$ is the error function defined as $\text {erf}(x)=\frac{2}{\sqrt\pi}\int_0^xe^{-t^2}dt$.
Making use of the previous estimators, the quantum channel parameters, i.e., the transmission $T$ and excess noise $\varepsilon$, can be estimated by
\begin{equation}
\label{eq13}
T=\frac{\hat t^2}{\eta},\varepsilon=\frac{\hat\sigma^2-N_0-\nu_{\text{el}}N_0}{{N_0}\hat t^2}.
\end{equation}

According to the analysis in Sec. \ref{sec2}, the transmitted Gaussian-modulated coherent states under the effects of reduced optical attenuation are different from the ideal states with perfect optical attenuation for a practical CVQKD system. It should be noted that $x^\prime_{A_0}$ and $x^\prime_B$ also satisfy with Eqs. (\ref{eq7}) and (\ref{eq8}), i.e.,
$x^\prime_B=tx^\prime_{A_0}+z$,
$\langle ({x^\prime_{A_0}})^2 \rangle=V^\prime_{X_{A_0}}, \langle{x^\prime_{A_0}}{x^\prime_B}\rangle=\sqrt{\eta T}V^\prime_{X_{A_0}}$,
$\langle ({x^\prime_B})^2 \rangle=\eta T V^\prime_{X_{A_0}}+\eta T \xi+N_0+V_{\text{el}}$.
Here, $V^\prime_{X_{A_0}}=V^\prime_{A_0}{N_0}$, $x^\prime_B$ is the practical value detected by Bob under the effects of imperfect optical attenuation.

However, Alice and Bob still use $x_{A_0}$ to estimate parameters for a CVQKD system with reduced attenuation under the condition that they do not aware the shifting of the attenuation level of VOA. Therefore, the relations between vectors kept by Alice and Bob become
\begin{equation}
\label{eq14}
\begin{split}
&\langle {x^2_{A_0}} \rangle=V_{X_{A_0}}, \langle{x_{A_0}}{x^\prime_B}\rangle=\sqrt{k\eta T}V_{X_{A_0}},\\
&\langle ({x^\prime_B})^2 \rangle=k\eta T V_{X_{A_0}}+\eta T \xi+N_0+V_{\text{el}}.
\end{split}
\end{equation}
Finally, based on Eqs. (\ref{eq8}) and (\ref{eq14}), we get
\begin{equation}
\label{eq15}
\begin{split}
&\sqrt{\eta T^\prime}V_{X_{A_0}}=\sqrt{k\eta T}V_{X_{A_0}},\\
&\eta T^\prime V_{X_{A_0}}+\eta T^\prime \xi^\prime+N_0+V_{\text{el}}=k\eta T V_{X_{A_0}}\\
&+\eta T \xi+N_0+V_{\text{el}}.
\end{split}
\end{equation}
Solving these equations yields
\begin{equation}
\label{eq16}
T^\prime=kT,\xi^\prime=\frac{\xi}{k}.
\end{equation}
Expressed in shot-noise unit, the estimated excess noise become
\begin{equation}
\label{eq17}
\varepsilon^\prime=\frac{\varepsilon}{k}(k>1).
\end{equation}

\begin{figure}[!h]\center
\centering
\resizebox{7.6cm}{!}{
\includegraphics{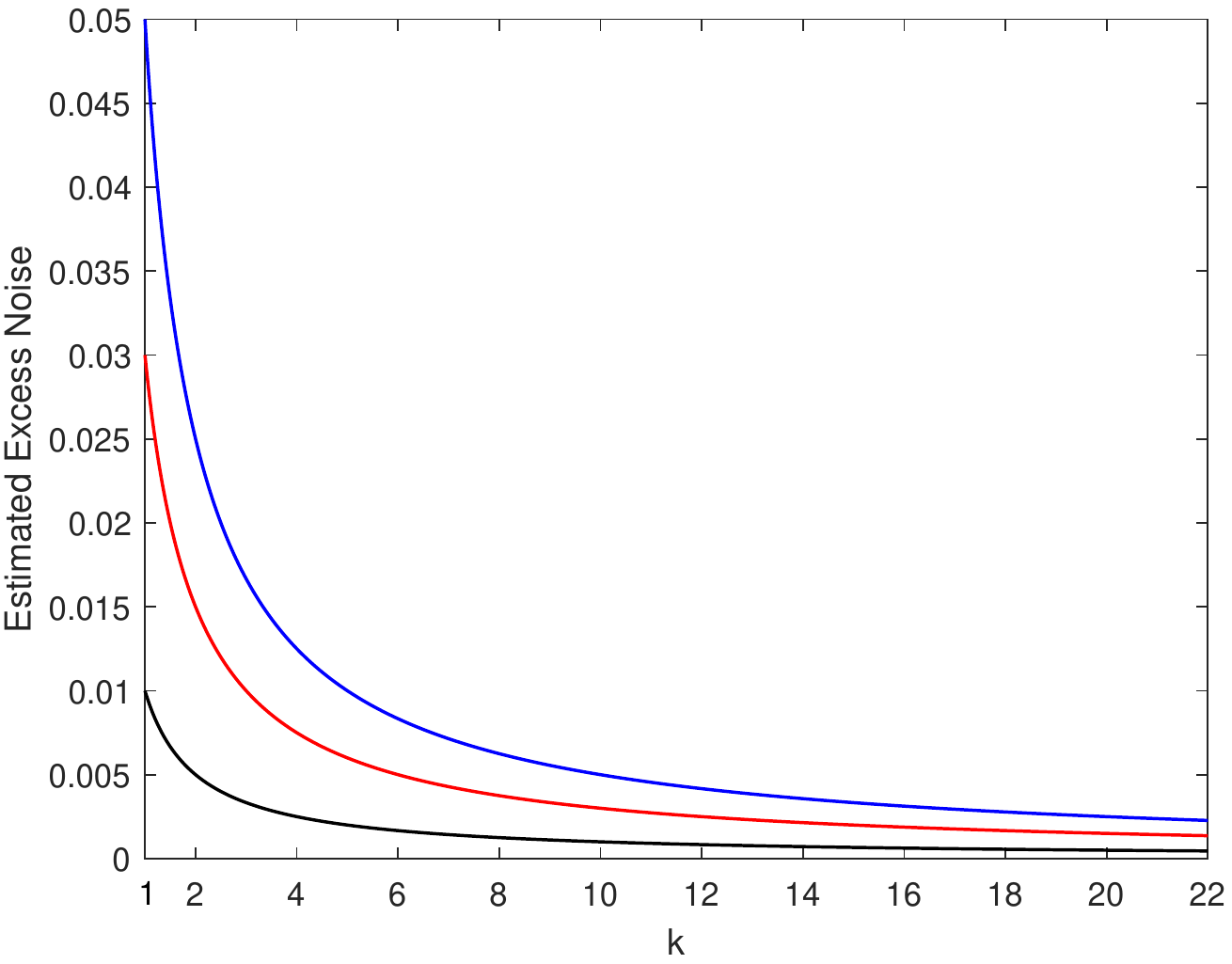}}
\caption{The estimated excess noise as a function of $k$ under different channel excess noise. Curves from top to bottom represent $\varepsilon=0.05$, $\varepsilon=0.03$, $\varepsilon=0.01$.}
\label{FIG6}
\end{figure}

Fig. \ref{FIG6} shows the estimated excess noise as a function of $k$ under $\varepsilon=0.01, 0.03, 0.05$, respectively. It is clearly that the channel excess noises are underestimated under the effects of reduced optical attenuation. We find that with the increment of the value of $k$, the estimated values of various channel excess noises will decrease. When the attenuation level deteriorates to a certain extent, the excess noise estimated by Alice and Bob may be close to zero. These analysis results indicate that the reduced optical attenuation may create a condition for Eve to conceal her attacks in a practical CVQKD system. Next, we use the classical partial intercept-resend (PIR) attack as an example to analyse the security of a practical CVQKD system under the effects of reduced optical attenuation.

In the PIR attack, the probability distribution of Bob's measurements is weighted sum of two Gaussian distributions \cite{Jouguet2013Preventing,lodewyck2007experimental}. The first one is the distribution of the intercepted resend data with a weight of $u$, and the other one is the distribution of the transmitted data with a weight of $1-u$. Further, the extra excess noise induced by the PIR attack is expressed by $2uN_0$. In theory, the excess noise estimated by Alice and Bob under the effects of the PIR attack should be written as
\begin{equation}
\label{eq18}
\xi_{\text{PIR}}=\xi_t+2uN_0,
\end{equation}
where $\xi_t=\varepsilon_t N_0$ is the technical excess noise of the system. Expressed in shot-noise, the estimated excess noise $\xi_{\text{PIR}}$ can be computed as
\begin{equation}
\label{eq19}
\varepsilon_{\text{PIR}}=\varepsilon_t+2u.
\end{equation}
Here, we use $u=0.2$ as an example to analyse the PIR attack in general situation. Correspondingly, the estimated excess noise become $\varepsilon_t+0.4$. In this case, the estimated excess noise under the effects of reduced optical attenuation should be calculated as
\begin{equation}
\label{eq20}
\varepsilon^\prime_{\text{PIR}}=\frac{\varepsilon_t+0.4}{k}.
\end{equation}

In a practical CVQKD system, we assume that the technical excess noise is a typical value, i.e., $\varepsilon_t=0.1$. Therefore, when Eve executes the PIR attack, the estimated excess noise under the effects of reduced optical attenuation can be calculated as $\varepsilon^\prime_{\text{PIR}}=\frac{0.5}{k}$. In the case of perfect optical attenuation, the noise value is $0.5$. The obvious increase of the excess noise makes the potential attack exposed. Hence, denial-of-service can occur to guarantee the security of the system. However, we find that Eve can exploit the effects of decreased optical attenuation to reduce the estimated excess noise. For example, when $k=5$, the estimated excess noise $\varepsilon^\prime_{\text{PIR}}=0.1$, i.e., the ideal noise value without attack. Here, based on Eqs. (\ref{eq1}) and (\ref{eq5}), Eve can reduce the attenuation level of the VOA by $10\lg5\approx7\text{dB}$ to satisfy with $k=5$. These analysis results fully demonstrate that the excess noise induced by the PIR attack can be completely hidden with the help of reduced optical attenuation. In particular, a full intercept-resend (FIR) attack can be implemented when $u=1$. Correspondingly, the excess noise estimated by Alice and Bob under the effects of reduced optical attenuation should be expressed as $\varepsilon_{\text{FIR}}=\frac{2.1}{k}$. In this case, when $k$ exceeds 21, the strongest FIR attack can be completely hidden by Eve.  Therefore, the decreased optical attenuation will open a loophole for Eve to hide her attacks, which seriously destroys the security of the practical system.
\section{SECRET KEY RATE AGAINST COLLECTIVE ATTACKS UNDER THE EFFECTS OF REDUCED OPTICAL ATTENUATION}\label{sec4}
The secret key rate and the maximal transmission distance are the key standards for the performance evaluation of a practical CVQKD system. Based on the parameters ${V_{A_0}}$, $T$, $\varepsilon$, $\eta$ and ${\nu _{\text{el}}}$, Alice and Bob can calculate the information shared by them, as well as the maximal bound on the information available to eavesdropper. In the case of reverse reconciliation, the secret key rate $K$ with $n$ received pulses used for key establishment against collective attacks is expressed as \cite{fossier2009improvement,leverrier2010finite}
\begin{equation}
\label{eq21}
K = \frac{n}{N}[ {\beta {I_{AB}} - S_{\text{BE}}^{\epsilon _{\text{PE}}} - \Delta \left( n \right)}],
\end{equation}
where $n = N - m$, and $\beta\in({0,1})$ is the reconciliation efficiency. $S_{\text{BE}}^{\epsilon_{\text{PE}}}$ represents the maximal value of the Holevo information compatible with the statistics except with probability ${\epsilon_{\text{PE}}}$, and ${I_{AB}}$ represents the shannon mutual information between Alice and Bob, which can be derived from Bob's measured variance ${V_B}$ and the conditional variance ${V_{B|A_0}}$ as
\begin{equation}
\label{eq22}
{I_{AB}} = \frac{1}{2}{\log _2}\frac{{{V_B}}}{{{V_{B| A_0}}}}=\frac{1}{2}{\log _2}\frac{{V_{A_0}}+1+{\chi_{\text{tot}}}}{1+{\chi_{\text{tot}}}}.
\end{equation}
Here, $\chi_{\text{tot}}=\chi_{\text{line}}+\chi_{\text{hom}}/T$ represents the total noise referred to the channel input, where $\chi_{\text{line}}=1/T-1+\varepsilon$, and $\chi_{\text{hom}}=[({1-\eta})+\nu_{\text{el}}]/\eta$. In particular, $S_{\text{BE}}^{\epsilon _{\text{PE}}}$ is determined by the following covariance matrix between Alice and Bob with finite-size effect:
\begin{equation}
\label{eq23}
\begin{split}
&{\Gamma _{AB}} = \\
&\left[ {\begin{array}{*{20}{c}}
{\left( {V_{A_0}} + 1 \right)\mathbb {I}}&{\sqrt {{T_{\min }}\left( {{V^2_{A_0}} + 2{V_{A_0}}} \right)} {\sigma _z}}\\
{\sqrt {{T_{\min }}\left( {{V^2_{A_0}} + 2{V_{A_0}}} \right)} {\sigma _z}}&{\left[ {{T_{\min }}\left( {{V_{A_0}} + {\varepsilon _{\max }}} \right) + 1} \right]\mathbb{I}}
\end{array}} \right],
\end{split}
\end{equation}
where matrices ${\mathbb {I}} = \left[ {\begin{array}{*{20}{c}}
1&0\\
0&1
\end{array}} \right]$ and ${\sigma _z} = \left[ {\begin{array}{*{20}{c}}
1&0\\
0&{ - 1}
\end{array}} \right]$, ${T_{\min }}$ and ${\varepsilon _{\max }}$ correspond the lower bound of $T$ and the upper bound of $\varepsilon$, respectively. According to Ref. \cite{leverrier2010finite}, when $m$ is large enough (e.g., $m>10^6$), $T_{\min}$ and $\varepsilon_{\max}$ can be calculated by using
\begin{equation}
\label{eq24}
\begin{split}
&T_{\min}=\frac{(\hat t-\Delta t)^2}{\eta},\\
&\varepsilon _{\max }=\frac{\hat\sigma^2+\Delta\sigma^2-N_0-\nu_{\text{el}}N_0}{\hat t^2N_0},
\end{split}
\end{equation}
where $\Delta t$ and $\Delta \sigma^2$ are defined in Eq. (\ref{eq12}). Then, $S_{\text{BE}}^{\epsilon_{\text{PE}}}$ can be acquired by
\begin{equation}
\label{eq25}
S_{\text{BE}}^{{ \epsilon _{\text{PE}}}} = \sum\limits_{i = 1}^2 {G\left( {\frac{{{\lambda _i} - 1}}{2}} \right)}  - \sum\limits_{i = 3}^5 {G\left( {\frac{{{\lambda _i} - 1}}{2}} \right)},
\end{equation}
where $G\left( x \right) = \left( {x + 1} \right){\log _2}\left( {x + 1} \right) - x{\log _2}x$, ${\lambda _i} \ge 1$ are sympletic eigenvalues derived from covariance matrices, which can be written as
\begin{equation}
\label{eq26}
\begin{split}
\lambda^2_{1,2}=&\frac{1}{2}(A\pm\sqrt{A^2-4B}),\\
\lambda^2_{3,4}=&\frac{1}{2}(C\pm\sqrt{C^2-4D}),\\
\lambda_5=&1,
\end{split}
\end{equation}
where
\begin{equation}
\label{eq27}
\begin{split}
A&=(V_{A_0}+1)^2-2T_{\text{min}}(V^2_{A_0}+2V_{A_0})\\
&+[T_{\text{min}}(V_{A_0}+\varepsilon _{\max })+1]^2,\\
B&=[(T_{\text{min}}\varepsilon _{\max }+1)(V_{A_0}+1)-T_{\text{min}}V_{A_0}]^2,\\
C&=\frac{A\chi_{\text{hom}}+(V_{A_0}+1)\sqrt B +T_{\text{min}}(V_{A_0}+\varepsilon _{\max })+1}{\eta T_{\text{min}}(V_{A_0}+\varepsilon _{\max })+1+\nu_{\text{el}}},\\
D&=\frac{\sqrt B(V_{A_0}+1)+B\chi_{\text{hom}}}{\eta T_{\text{min}}(V_{A_0}+\varepsilon _{\max })+1+\nu_{\text{el}}}.
\end{split}
\end{equation}

Moreover, $\Delta \left( n \right)$ is a linear function of $n$ in Eq. (\ref{eq21}), which is related to the security of the privacy amplification. In a practical CVQKD system, it can be given by \cite{leverrier2010finite}
\begin{equation}
\label{eq28}
\Delta \left( n \right) = 7\sqrt {\frac{{{{\log }_2}\left( {{1 \mathord{\left/
 {\vphantom {1 {\overline  \epsilon  }}} \right.
 \kern-\nulldelimiterspace} {\overline  \epsilon  }}} \right)}}{n}}  + \frac{2}{n}{\log _2}\frac{1}{{{ \epsilon _{\text{PA}}}}},
\end{equation}
where $\overline \epsilon$ and ${ \epsilon _{\text{PA}}}$, which are virtual parameters and can be optimized in the computation, denote the smoothing parameter and the failure probability of the privacy amplification, respectively. In addition, $\overline  \epsilon$ and ${ \epsilon _{\text{PA}}}$ are usually set to be equal to ${ \epsilon _{\text{PE}}}$ due to the value of $\Delta (n)$ mainly depends on $n$.

Based on the above equations, Alice and Bob can calculate the secret key rate with finite-size effect against collective attacks for the system. Therefore, the secret key rate can be regarded as a function of the above parameters, i.e., $K=K(V_{A_0}, T, \varepsilon, \nu_{\text{el}})$. According to the analysis in Sec. \ref{sec3}, the estimated quantum channel parameters under reduced optical attenuation are different from their true values. In addition, the variation of $V_{A_0}$ can also not be detected by Alice and Bob when the modulation variance is not be monitored in a practical CVQKD system. Accordingly, the evaluated secret key rate for a CVQKD system with decreased optical attenuation is expressed as $K_e=K(V_{A_0}, T^\prime, \varepsilon^\prime, \nu_{\text{el}})$. However, the practical secret key rate should be calculated as $K_p=K(V^\prime_{A_0}, T, \varepsilon, \nu_{\text{el}})$. In order to clearly reveal the difference between the evaluated secret key rate and the practical secret key rate for a practical CVQKD system with reduced optical attenuation, we simulate the secret key rate versus transmission distance under different excess noise $\varepsilon$.

\begin{figure}[!h]\center
\centering
\resizebox{7.6cm}{!}{
\includegraphics{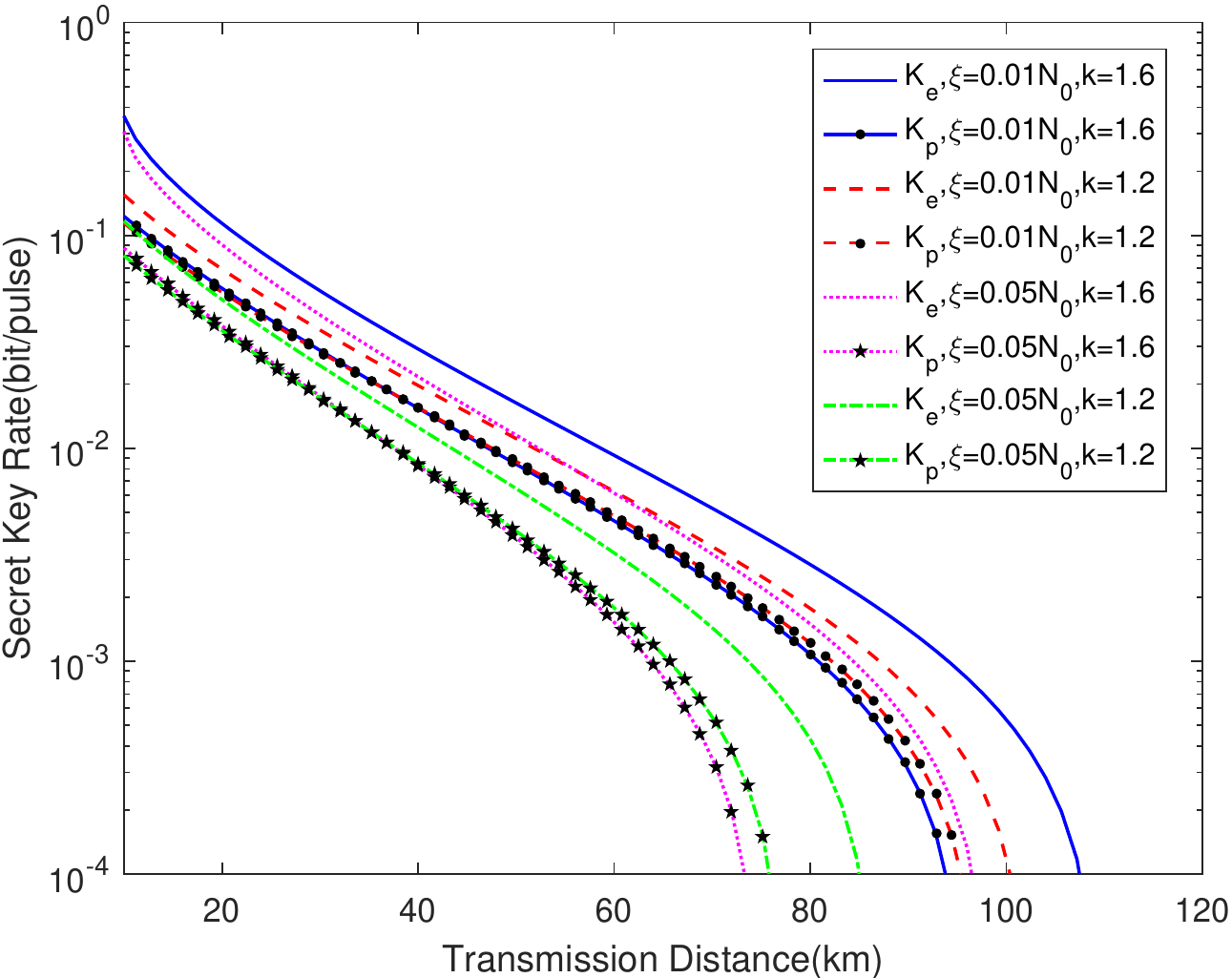}}
\caption{Secret key rate vs transmission distance for different deterioration situations when $\varepsilon=0.01,0.05$, respectively. Unlabeled curves from top to bottom represent the relations between the evaluated secret key rate $K_e$ and the transmission distance under different imperfect attenuation situations. Labeled curves from top to bottom show the corresponding practical secret key rate $K_p$ vs transmission distance under different deterioration situations. The fiber loss is 0.2 $\text{dB}/\text{km}$.}
\label{FIG7}
\end{figure}

Fig. \ref{FIG7} depicts the relationship between the secret key rate and the transmission distance for a practical CVQKD system under the effects of different deterioration situations when $\varepsilon=0.01,0.05$. The fixed parameters for the simulation are set as: $V_{A_0}=4, \eta=0.5, \nu_{\text{el}}=0.01, \beta=95\%, \epsilon=10^{-10}, m=0.5\times N$, respectively. It is obvious that the secret key rate $K_e$ evaluated by Alice and Bob under the effects of reduced optical attenuation is overestimated compared with their practical value $K_p$. The simulation results also demonstrate that the decreased optical attenuation can open a security loophole for Eve to perform an intercept-resend attack in a practical CVQKD system.

In particular, the difference between the estimated secret key rate under reduced optical attenuation and the practical secret key rate in the same situation reflects the key information that can be acquired by Eve through the intercept-resend attack. We find that the leaking of the secret key information increases with the decrease of optical attenuation. In addition, the simulation results also indicate that Eve can acquire more secret key information in the case of a larger excess noise $\varepsilon$ under the effects of the same amount of decrease of optical attenuation.

More importantly, the defense of the attack is a key task. We observe that the changes of the modulation variance of the system and the attenuation level of VOA are synchronous in a practical CVQKD system. Therefore, the real-time monitoring of the modulation variance of the system might close this loophole, which is investigated in detail in next section.
\section{COUNTERMEASURES}\label{sec5}
The above investigations show that the influences of reduced optical attenuation effect on the estimated parameters procedures and the evaluated secret key rate. To remove the loophole induced by reduced optical attenuation, we here add an isolator at Alice's output port to resist the potential laser damage attack. More specifically, we first bound the power of Eve's injected light to a reasonable based on the method in Ref. \cite{lucamarini2015practical}, and then choose an appropriate isolator for the system. However, it might be possible that Eve is able to decrease the performance of the isolator by using the laser damage attack. Therefore, the isolation value should be monitored in real time. In order to reduce costs and improve efficiency, we can replace the isolator with an optical fuse. The optical fuse can only tolerate a certain amount of laser power but disconnects itself once the power crosses a threshold. In this way, the optical fuse physically blocks the injected high power, which effectively protects the system from the laser damage attack.

Apart from the laser damage attack, the attenuation value may decrease due to the effects of the rising of environment temperature. Accordingly, we propose a real-time monitoring scheme for the level of optical attenuation to prevent the incorrect estimation of channel parameters. According to the analysis of Section. \ref{sec2}, we find that the performance of the optical attenuator can be evaluated by monitoring the real-time variation of the modulation variance before the transmission of quantum signal.

\begin{figure}[!h]\center
\centering
\resizebox{8.5cm}{!}{
\includegraphics{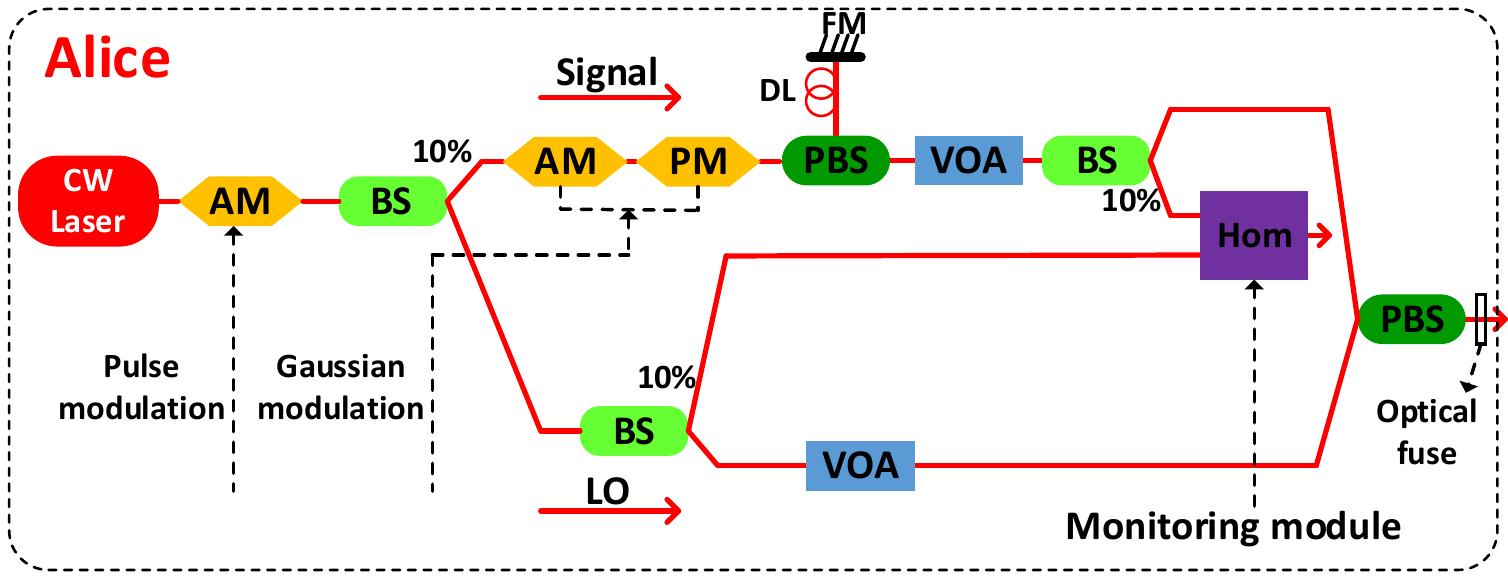}}
\caption{The structure of the real-time monitoring scheme for the level of optical attenuation in Alice's apparatus. }
\label{FIG8}
\end{figure}
Fig. \ref{FIG8} shows the procedure of the real-time monitoring scheme of the level of optical attenuation for a practical CVQKD system in Alice's apparatus. Specifically, Alice first splits a fraction of the attenuated quantum signal and the undamped LO signal. Then, Alice can get a sample $U$ after the interference between the separated quantum signal and LO signal in a homodyne detector. The detection module has been described in Fig. \ref{FIG4}. In the sample, all data are the sampled voltage values. Then, the variance of these data can be counted as
\begin{equation}
\label{eq29}
\begin{split}
\text{var}(U)&=\langle U^2 \rangle-{\langle U \rangle}^2\\
&=\frac{1}{N_u}\sum_{i=1}^{N_u} U^2_i-(\frac{1}{N_u}\sum_{i=1}^{N_u}U_i)^2,
\end{split}
\end{equation}
where $N_u$ is the size of the sample, $U_i$ is the ith data in the sample. Since the sample is finite, the value of $\text{var}(U)$ should be modified as
\begin{equation}
\label{eq30}
\text{var}^\prime(U)=\text{var}(U)+z_{\epsilon_{\text{PE}}/{2}}\frac{\text{var}(U)\sqrt2}{\sqrt{N_u}}.
\end{equation}
Subsequently, the measured variance of the quadrature variables $x_{A_0}$ or $p_{A_0}$ of the separated quantum signal can be calculated as \cite{laudenbach2018continuous}
\begin{equation}
\label{eq31}
V_m=\langle x^2_{A_0} \rangle_m= \langle p^2_{A_0} \rangle_m=\frac{\text{var}^\prime(U)}{P_{\text{LO}} \rho^2 g^2 Bhf},
\end{equation}
where $P_{\text{LO}}$ is the power of the separated local oscillator, $\rho$ is the PIN diode's responsivity, $g$ is the total amplification of the homodyne detector, $B$ is electronic bandwidth, $h$ is Planck's constant, $f$ is optical frequency. Because these parameters are some constant values, Alice can analyse the variation of the modulation variance through the sample $U$. Further, the practical variance of the quadrature variables $X_{A_0}$ or $P_{A_0}$ can be calculated as
\begin{equation}
\label{eq32}
V_p=V^\prime_{A_0}=\frac{V_m-N_0-\nu_{\text{el}}N_0}{N_0}.
\end{equation}
Therefore, $k$ can be acquired in real time by analysing the difference between $V_p$ and the preset modulation variance $V_{A_0}$, i.e., $k=\frac{V^\prime_{A_0}}{V_{A_0}}=\frac{V_p}{V_{A_0}}$. Finally, Alice and Bob can utilize the acquired $V_p$ and $k$ to precisely evaluate the secret key rate of the system, i.e., $K_m=K(V_p,\frac{T^\prime}{k},k\varepsilon^\prime,\nu_{\text{el}})=K_p$. These analysis results fully indicate that the real-time monitoring scheme can effectively resist the loophole induced by reduced optical attenuation.

Importantly, in the proposed scheme, the homodyne detector and the separated LO signal may be controlled by Eve. Fortunately, the countermeasures against the attacks related to LO and homodyne detector has been designed well in Refs .\cite{Wang2016Practical,Jouguet2013Preventing,Qin2016Quantum,qin2018homodyne}. Based on these corresponding countermeasures, the proposed monitoring scheme is immune to Eve. In addition, the monitoring module may reduce the performance of the system, and add the complexity of the implementation of the system. However, it is notable that the influences of the losses of the Gaussian-modulated signal and LO signal can be completely eliminated by properly adjusting the preset value of the attenuation level of VOA in a practical CVQKD system.

Through the combination of the above countermeasures, the modulation variance of the system can be precisely evaluated in real time. Eventually, the secret key rate of the CVQKD systems with reduced optical attenuation can be precisely evaluated to effectively remove this security loophole with the help of these schemes.
\section{CONCLUSION}\label{sec6}
We have investigated the impacts of reduced optical attenuation for a practical CVQKD system. We reveal that the transmitted Gaussian-modulated coherent states will deviate from the preset values under the effects of decreased optical attenuation. Further, we find that the estimated channel excess noise is smaller than the practical value under the influences of reduced optical attenuation. This effect makes the secret key rate of the system overestimated, which can open a security loophole for Eve to successfully perform the intercept-resend attack in a practical CVQKD system. We show that Eve can obtain more key information for the case of a larger channel excess noise in the same amount of decrease of optical attenuation. Eve also can obtain more key information by reducing the attenuation value further. In order to close the loophole induced by reduced optical attenuation, we first add an optical fuse at Alice's output port and then propose a real-time monitoring scheme for the attenuation level of VOA in a practical CVQKD system by analysing the variation of the variance of the quadrature variables $X_{A_0}$ or $P_{A_0}$ of the attenuated quantum signal before send to Bob. These countermeasures can make Alice and Bob precisely evaluate the channel parameters to accurately analyse the performance of the CVQKD systems with reduced optical attenuation.
\section*{ACKNOWLEDGMENTS}
This work was supported by the National key research and development program (Grant No. 2016YFA0302600), the National Natural Science Foundation of China (Grants No. 61671287, 61631014, 61332019, 61632021), and the Nature Fund of Science in Shaanxi in 2018(Grant No. 2018JM6123).

\end{document}